\documentclass[useAMS,usenatbib]{mn2e}
\usepackage{graphicx}

\title[Comparison of Debrecen and Greenwich sunspot areas]{Indirect comparison of Debrecen and Greenwich daily sums of sunspot areas}
\author[T. Baranyi et al.]{T.~Baranyi,$^1$\thanks{E-mail: baranyi.tunde@csfk.mta.hu}
  S.~Kir\'aly,$^2$ H.E.~Coffey,$^3$\\
  $^1$Heliophysical Observatory, Research Centre for Astronomy and Earth Sciences, Hungarian Academy of Sciences, \\
  Debrecen, P.O.Box 30, H-4010, Hungary\\
  $^2$Konkoly Observatory, Research Centre for Astronomy and Earth Sciences, Hungarian Academy of Sciences,\\
  Budapest, Konkoly Thege Mikl\'os \'ut 15-17, H-1121, Hungary\\
  $^3$NOAA National Geophysical Data Center, 325 Broadway, Boulder, CO, 80305, USA}

\begin{document}

\date{Accepted 2013 June 19. Received 2013 June 17; in original form 2012 November 1}

\pagerange{\pageref{firstpage}--\pageref{lastpage}} \pubyear{2013}

\maketitle

\label{firstpage}

\begin{abstract}
Sunspot area data play an important role in the studies of solar activity and its long-term variations. In order to reveal real long-term solar variations precise homogeneous sunspot area databases should be used. However, the measured areas may be burdened with systematic deviations, which may vary in time. 
Thus, there is a need to investigate the long-term variation of sunspot area datasets and to determine the time-dependent cross-calibration factors.
In this study, we investigate the time-dependent differences between the available long-term sunspot databases. Using the results, we estimate the correction factor to calibrate the corrected daily sunspot areas of Debrecen Photoheliographic Data (DPD) to the same data of Greenwich Photoheliographic Results (GPR) by using the overlapping Kislovodsk and Pulkovo data. We give the correction factor as $GPR=1.08(\pm 0.11) *DPD$.

\end{abstract}


\section{Introduction}

The sunspots are the most easily observable manifestations of solar activity.
They are related to those regions of solar magnetic flux where the concentration of magnetic field is strong enough to lead to a lower temperature on the solar surface compared to its surroundings (Solanki 2003). 
The routine measurements of daily sunspot area data were started in 1874.
Now the sunspot area records provide the longest available time series among the physical indices of solar activity (Usoskin 2008). Thus, the time series of the real sunspot area (the area corrected for foreshortening) are important for different types of studies of solar activity.

The long-term studies need precise homogeneous databases, but the sunspot area data of different observatories may be burdened with systematic deviations. 
Consequently, different data sets must be combined after an appropriate cross-calibration to create a reliable time series of sunspot areas (Balmaceda et al. 2009).
The systematic differences between datasets are related to the different observational techniques and different data reduction methods (Baranyi et al. 2001). There are larger systematic deviations between the area databases derived from photographic observations and sunspot drawings  but there are also smaller differences within these two subsets.
However, the instruments and methods used in a given observatory may vary on a decadal or longer time scales, too. For example, the gradual spread of usage and increasing resolution of CCD cameras as well as the parallel decay of the quality of the available photographic plates during the last 10-15 years forced lots of observing sites to replace the plates with CCD camera, and following from this to change their measuring method/software. Those databases, which use observations of several sites may also vary because the fraction of contributors changes, or some sites cease the observations/measurements, or some new ground-based or space-borne observing sites join to the data providers.
The mentioned changes are unavoidable, thus it is almost impossible to keep any area dataset in a perfectly homogeneous state. On the one side, this makes their cross-calibration difficult. On the other side, this shows that the comparison of different datasets can be important even in that case when someone wants to use only one of the datasets in his/her studies. The time series of calibration factors can help to reveal how the studied variations of solar activity may be burdened with some intrinsic variations of that dataset.

The first sunspot database was published by the Royal Greenwich Observatory in the volumes of GPR between 1874 and 1976.
After 1976, the Debrecen Heliophysical Observatory took over the responsibility of publication of sunspot data for every day with the endorsement of IAU. There is no overlap between these two datasets at present, thus the cross-calibration between them can be made only in an indirect way by using independent overlapping databases. However, the result may also depend on the time-dependent variations of these databases.
The aim of this study is to reveal the time-dependent systematic differences between the available long-term sunspot databases, and after that to estimate the correction factor between GPR and DPD.

\section{Sunspot area databases}

The GPR catalog contains position and area data of sunspot groups measured in photographic observations taken at Royal Observatory of Greenwich or Cape of Good Hope, and at the Kodaikanal Observatory, and at a few other observatories. 
The photoheliograph of RGO was modified several times during the decades, and in May 1949, it was moved from Greenwich to Herstmonceux (Sussex) to have better observing conditions (McCrea 1975). Further details on GPR data and their digital versions have been published by Willis et al. (2013a) recently.
In this study, the digital version of GPR available at NOAA National Geophysical Data Center (NGDC)\footnote{http://www.ngdc.noaa.gov/stp/solar/solardataservices.html} is used.

   The DPD catalog (Gy\H ori, Baranyi \& Ludm\'any 2011) contains the same type of data as GPR for the whole group and each spot in it. The DPD data are measured on daily white-light full-disk photographic plates, which are mainly taken at Debrecen and its Gyula Observing Station with 10-11 cm solar diameter. 
The measuring technique and the quality of the photographic plates available for observations have changed a lot since the start of measurements, but the largest part of DPD data are reduced from high contrast films, digitized with about 8000x8000 pixels spatial resolution at 16-bit gray scale levels. The recent changes in the Debrecen-Gyula observations are that the cassette of film at Gyula was replaced with a CCD camera of 4000x4000 pixels in 2009, which was replaced with a camera of 8000x6000 pixels in 2011.  
Any gaps are filled in with the observations of cooperating ground-based  observatories, or in some cases, the data derived from space-borne SOHO/MDI quasi-continuum images. The number of the contributing sites and the number of their observations measured for DPD also vary in time. Thus, the systematic differences because of their different observational material and spatial resolution may also cause some time-dependent variations in DPD data.
A special characteristic of DPD is that the volumes of consecutive years of sunspot area data were measured, not in fully chronological order but as resources, and data availability permitted. This decreases any possible chronological order bias where some systematic differences due to changes of the measuring method may be propagated as time progresses. For example, the starting and ending years of the studied interval have been measured at about the same time recently, thus the change of the measuring method can hardly cause any differences between these years. The whole DPD dataset is still regularly revised and improved\footnote{http://fenyi.solarobs.unideb.hu}.

   Between 1932 and 1991 {\it Catalogues of solar activity} ("Katalogi solnechnoj deyatel'nosti") were published in Central Astronomical Observatory at Pulkovo. In this catalogue the subsection "Sunspot areas" of Section "Daily indices" contains daily sums of total corrected sunspot areas, which have been derived from observations taken in the former USSR. The electronic database of these data is available at the site of the Pulkovo Observatory\footnote{http://www.gao.spb.ru/database/csa/main\_e.html}. In this study, we will refer to this database as Pulkovo. In a not perfectly independent way, publication of solar data in the monthly bulletin {\it Solar Data} ("Solnechnie Dannie") was started at Pulkovo in 1954. The data were mainly based on the  photographic plates taken at Pulkovo and its Kislovodsk Mountain Astronomical 
Station, but this dataset also contains data derived from the observations of the unified Soviet solar network and other cooperating observatories. 
This database is now available as {\it Kislovodsk sunspot group reports}, thus we will refer to it as Kislovodsk\footnote{http://www.solarstation.ru/}.  The Kislovodsk data are particularly suitable for extending the series of sunspot areas because this project has a stable technical and personnel background (Nagovitsyn, Makarova \& Nagovitsyna 2007). The Kislovodsk and Pulkovo data can be reckoned as quasi-extensions of each other because of the strong connection between them. 
In addition to publishing its own data, Kislovodsk also contributes to DPD. The days with Kislovodsk observations cover about 7\% of the whole interval of DPD, but the plates observed at Kislovodsk are measured in Debrecen independently from the Kislovodsk data. Thus, these two datasets can be reckoned as independent time-series of sunspot area. The t-test has confirmed that there is no significant difference between the means of correction factors derived from DPD including or excluding Kislovodsk data (the mean including Kislovodsk data is 0.925; the mean excluding Kislovodsk data is 0.922; the significance (2-tailed) value is 0.797).

   At the Rome Observatory, three different telescopes were used over the whole period of measurements (1958-1999). All sunspot measurements were performed on photographic enlargements of the original photographic observation. 
The measurements were performed on a solar disk of approx 15 cm diameter with Stonyhurst disks, by following common rules (Cimino 1967; Torelli \& Ermolli 2009 personal communication). The data were published in the volumes of {\it Solar Phenomena Monthly Bulletin}, which contain two tables for the area data. 
The first table contains the daily sums of whole area of groups and the second one contains area data of the individual groups. 
Now the NGDC as WDC hosts the Rome data at its site\footnotemark[1].
At Rome sometimes some sunspot groups are omitted from the area measurements, and only their position data are published. This important information was indicated in different ways during the decades (typed in Italic, marked with asterisk or other type of symbol) in the section of daily data. Sometimes this information was missing from the digitized version of the dataset. Thus, we checked the set of marks in the files of the daily data, and we also revised it by using the files of group data.
Only the daily sums without missing data are involved in this study. 

The solar patrol at Boulder Solar Observatory started in 1966. The observatory became one of the NASA SPAN (Solar Particle Alert Network) sites in 1967, a world network of solar stations established to provide warnings of hazardous solar particle events.  The Solar Optical Observing Network (SOON) of the US Air Force took over this task from the SPAN network in the 1970s, and  Boulder operated as a joint NOAA/USAF facility.  
A 4-inch refractor was used to make drawings of 20 cm original diameter, which are available for 1966-1992 at the site of NGDC together with the files of digitized sunspot region data for 1966-1981. We used these scanned observations to quality-check the files of group area and after that we computed the daily sums for 1966-1981. The daily sums for 1982-1993 were computed by summarizing the Boulder group area from the USAF\_MWL reports\footnotemark[1]. 

The SOON network maintains 24-hour synoptic solar patrol producing real-time data, thus the quick data reduction is of the essence. Although detailed and extremely accurate measurements are not possible at the SOON stations, the SOON dataset became especially important and valuable. It was almost exclusively used during the last three decades when only a few sunspot datasets were available. Even today, the SOON data are frequently used in a number of studies, and a highly processed version of SOON data
are used routinely to update the GPR to the present\footnote{http://solarscience.msfc.nasa.gov/greenwch.shtml}. The DPD sunspot group numbering is also based on the NOAA sunspot group numbers of SOON network.
The 1-year sunspot region tables are compiled from the SOON reports at NOAA NGDC, and after some quality check of data, final reports are published including all SOON stations data in the USAF\_MWL reports. These final reports and the sunspot drawings of 18 cm diameter are available at the site of NGDC\footnotemark[1]. 
The scaling of positions and areas are done routinely by hand, using Stonyhurst disc overlays for both elements. The data were reduced by USAF staff who rotated into their solar technician jobs for two years and then in many cases moved on to other jobs. The area data contain some estimation because "one may well find a greater quantisation of reported areas, because of the finite number of circles and ellipses on the overlay, although some observers will attempt to estimate a value in between the template areas" (Kennewell 2004\footnote{ftp://ftp.ngdc.noaa.gov/STP/SOLAR\_DATA/SUNSPOT\\\_REGIONS/USAF\_MWL/docs/SunspotAreaMetadata.txt}). Due to the sometimes extreme variability of the SOON sunspot areas, NGDC chose to publish all the stations data and let the user community decide which station's values were more accurate.  By the 1990s, NGDC was eliminating extreme outliers.
The number of the sites and the members of the network varied in time. 
In this study the SOON sites with the longest datasets are included: 
Holloman, Learmonth, San Vito, and Ramey.

The datasets studied in this paper are listed in Table 1. The daily sums are selected from those databases which contain these sums (GPR, DPD, Pulkovo, and Rome), or they are computed from the daily group data if only those are available (Kislovodsk, Boulder, and the SOON sites).
\begin{table}
 \centering
  \caption{Observational data.}
  \begin{tabular}{llll}
  \hline
   Database     &    Studied years        &  Type of      &Min. corrected \\
                &                         &  observation  &area reported \\
 \hline
 GPR & 1933-1976 & photographic & 1 msh \\
 Pulkovo & 1933-1991 & photographic & 1 msh \\
 Kislovodsk & 1955-2011 & photographic & 1 msh \\
 Rome & 1958-1999 & photographic & 2 msh \\
 DPD & 1977-2011 & photographic & 1 msh \\
 Boulder & 1966-1993 & drawing & 10 msh \\
 Holloman & 1982-2010 & drawing & 10 msh \\
 Learmonth  & 1982-2010 & drawing & 10 msh \\
 Ramey  & 1982-2002 & drawing & 10 msh \\
 San Vito  & 1987-2010 & drawing & 10 msh \\
\hline
\end{tabular}
\end{table}

\section{Method of analysis}

To determine the cross-calibration factors, we follow the main steps of the method of Balmaceda et al. (2009) with a few modifications  in the selection criteria of the data excluded from the study. Ordinary least-squares (OLS) linear regression model without an intercept term is applied to the data of two different observatories ($X$ and $Y$), i.e. the slope of a linear regression forced to pass through the origin:
\begin{equation}
   Y=b*X.
\end{equation}

   In the first analysis, the model is applied to all the points when both of the pair of data are larger than zero. The zero values are excluded 
because they do not refer to a real relationship between area measurements. 
The slope derived in this way is taken to be the initial estimate for a second analysis, and the standard error of the estimate ($\sigma_{fit}$) is determined. In the second analysis, the zero values are excluded again but outliers are also excluded when only points within $3*\sigma_{fit}$ from the first fit are taken.

  To reduce the impact of measurement errors, we have to treat the variables symmetrically. Thus, we apply an inverse linear regression, switching the dependent variable with the independent variable. We follow the method of OLS-mean (Babu \& Feigelson 1992) which takes the arithmetic mean of the two OLS slopes as an estimate of the slope of the regression line. This method works quite well when the correlation coefficient is larger than 0.5. This criteria is fulfilled in our case as the correlation coefficient is higher than 0.95 in each case.

The procedure described in the previous paragraph is repeated in the inverse linear regression:
\begin{equation}
   X=b'*Y.
\end{equation}

 The correction factor for $X$ is the mean slope calculated as $1/2*(b+1/b')$. 
 This correction factor is used to adjust the average level of a dataset $X$ to that of $Y$. The standard error of the mean slope ($\sigma_{slope}$) is calculated as 
 $\sigma_{slope}=\sqrt{1/4*[Var(b)+Var(1/b')+2*Cov(b,1/b')]}$ by using the variances and covariance of the slopes of $OLS(Y | X)$ and $OLS(X | Y)$ according to the related formula in Table I 
of Babu \& Feigelson (1992).

\begin{figure*}

\centering
\includegraphics[width=14cm]{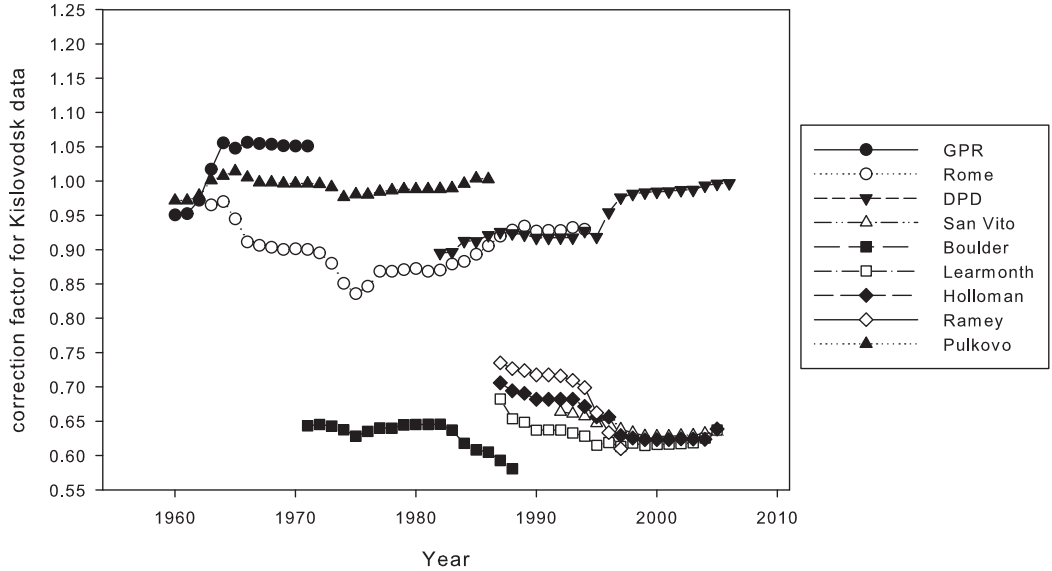}

 \caption{The time series of the 11-year correction factors for Kislovodsk daily sums of corrected sunspot area data to calibrate them to the other datasets computed over an 11-year sliding window. The symbols are at the centers of the related 11-year intervals. }

 \label{fig1}

\end{figure*}
\begin{figure*}

\centering
\includegraphics[width=14cm]{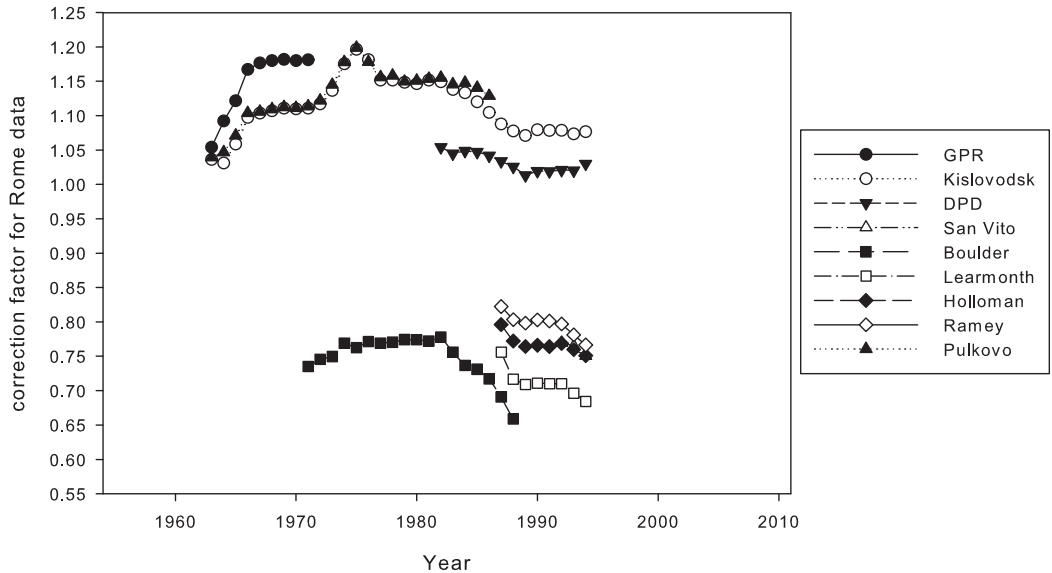}

 \caption{The same as Figure 1 but for Rome sunspot area data.}

 \label{fig2}

\end{figure*}

\section{COMPARISON OF DATABASES}

In this study, the daily sums of the corrected area in millionths of a solar hemisphere are used, and two kinds of correction factors are calculated based on them. The first type is calculated by using an 11-year interval of daily sums, an extensive interval which gives a large enough statistical sample for a comparison. This is called 11-year correction factor. The second type is calculated in a 1-year interval of daily sums, which is called 1-year correction factor. This shows larger variations because of the smaller sample time length, but it easily shows the short-time changes of the databases. 

At first, we calculate the 11-year correction factors within an 11-year sliding window. In Figure 1 the independent variable is Kislovodsk while in Figure 2 it is Rome. In these figures we can study the time dependence of the 11-year correction factors based on the two longest independent overlapping datasets. 
 The dependent variables are indicated in the box of legends. The standard error of the slope ($\sigma_{slope}$) is about 0.005 in each case.

\begin{figure*}

\centering
\includegraphics[width=14cm]{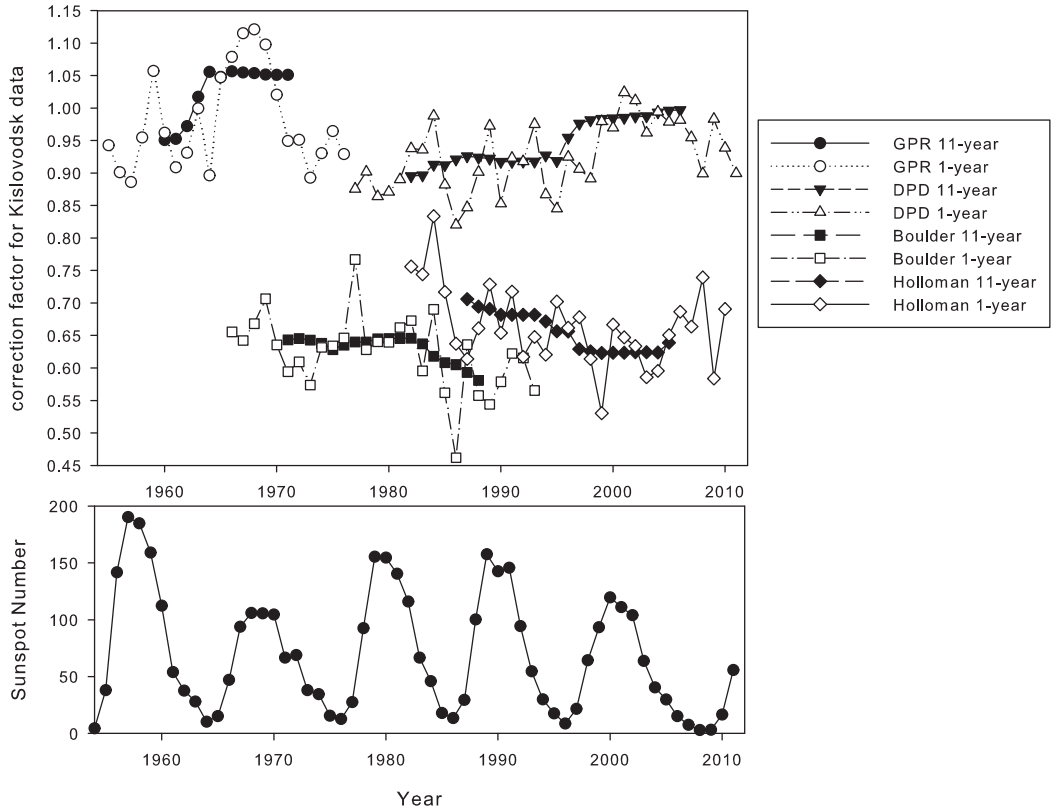}

 \caption{Upper panel: The 11-year correction factors for Kislovodsk daily sums are plotted using four of the stations as shown in Figure 1, and the related 1-year correction factors are added to show their relationships. Lower panel: Yearly mean of the International Sunspot Number.}

 \label{fig3}

\end{figure*}

\begin{figure*}

\centering
\includegraphics[width=12.17cm]{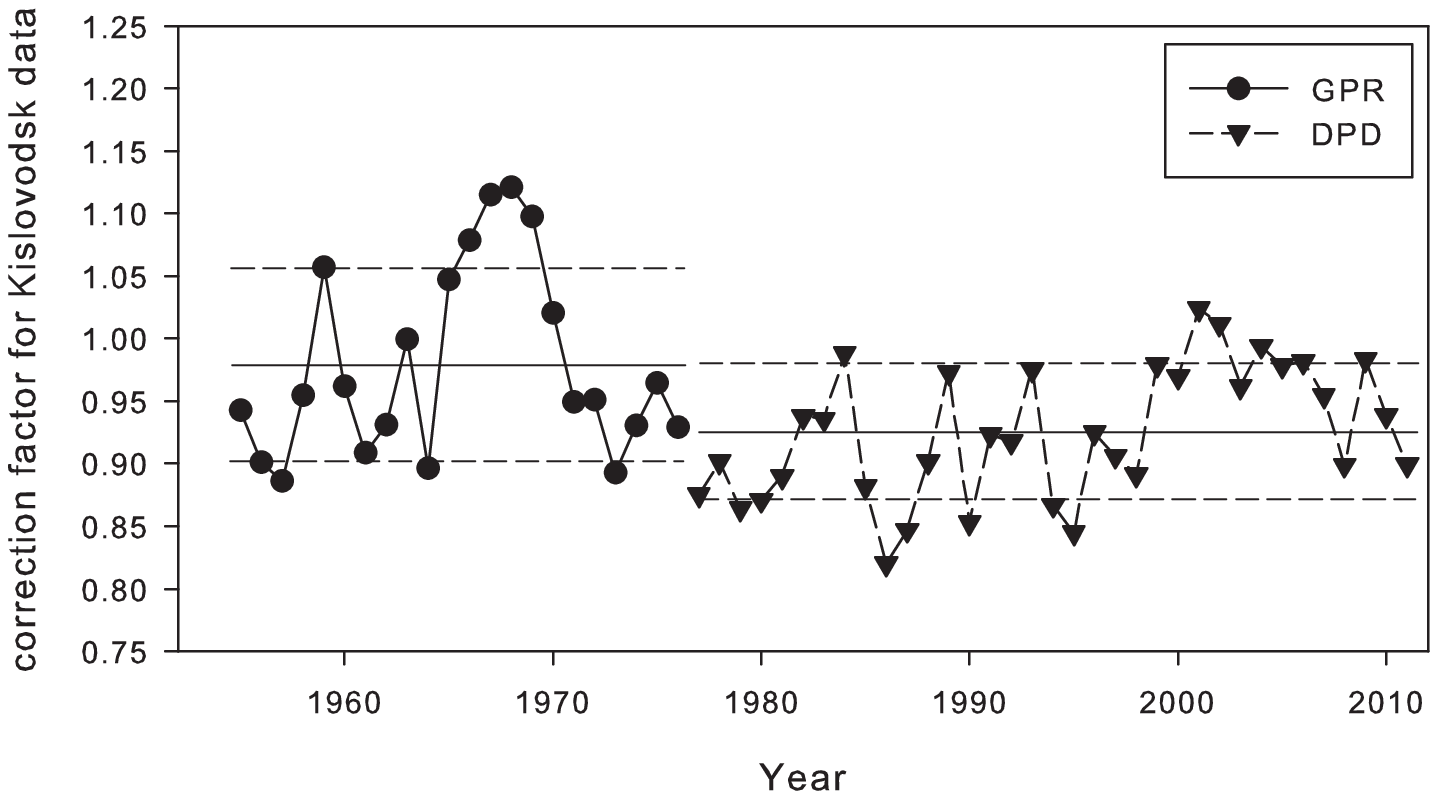}

 \caption{The 1-year correction factors for Kislovodsk daily sums for the years 1955-2011 to calibrate them to the other datasets. The horizontal solid lines show the means of the related 1-year correction factors, and the horizontal dashed lines show the the error bands determined by the Equations 3 and 4.}

 \label{fig4}

\end{figure*}

\begin{figure*}

\centering
\includegraphics[width=11cm]{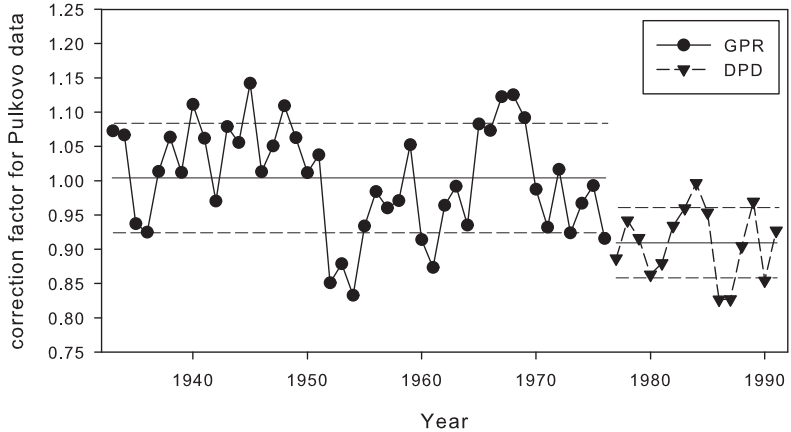}

 \caption{The 1-year correction factors for Pulkovo daily sums for the years 1933-1991 to calibrate them to the other datasets. The horizontal solid lines show the means of the related 1-year correction factors, and the horizontal dashed lines show the error bands determined by the Equations 9 and 10. }

 \label{fig5}

\end{figure*}

The most conspicuous feature is that every 11-year correction factor varies in time in both figures. 
The amplitude of the long-term variation is about 10-20 \% in each case. The variation is not persistent; there are short intervals when the relationship between the dependent and independent variables are stable. When the correction factor changes, it is difficult to determine which dataset has an internal variation of its average level: the dependent variable, or the independent variable, or both of them in a different extent.
The comparison of Figures 1 and 2 does not help in the decision. We have no enough data to decide when and how GPR, Kislovodsk, Rome or DPD vary. 

To estimate the uncertainties of the 11-year correction factors, we have to use the 1-year correction factors too.  The relationship between the 11-year and 1-year correction factors can be seen in Figure 3 including four cases as sample cases. This figure shows that a better estimation of the uncertainty of a 11-year correction factor can be computed by combining the variance of the 1-year correction factors and the variance of the slope. This means that the uncertainties would be about between $\pm 0.03$ and $\pm 0.1$ if we used the 11-year correction factors to determine the final correction factor. 
However, the 11-year correction factor is not the best choice for this estimation because it shows somewhat larger sensitivity to the variations around maximum years than around solar minima as Figure 3 show. The reason for this behaviour is that the variance of the sunspot area data increases with the increase of area. When the center of the 11-year sliding window moves forward in time to the next year, only 365 (or 366) daily data are replaced with new ones. As data arrive sequentially in the time series database, the sliding window of 11 years deletes the data of the first year in the window and adds new daily data at the end. The change of the 11-year correction factor depends on the statistical difference between these two years. The differences between the maximum years of two consecutive cycles can be much larger in absolute value than the differences between their minimum years. Thus, a one-year shift of the sliding window can cause a larger change of the slope of the regression line when a maximum year (from the beginning of the interval) is replaced with a maximum year of the next cycle (at the end of the interval) than in the case of the exchange of minimum years. The Figure 3 also shows that the actual value of the correction factor strongly depends on the length and the starting year of the interval used for the calculation because of the time-dependent deviations.

Thus, to estimate the final correction factor between GPR ($G$) and DPD ($D$), we use the 1-year correction factors in the longest available overlapping intervals. In our case, Kislovodsk ($K$) and Pulkovo ($P$) have the longest overlaps with both GPR and DPD. The 1-year correction factors for $K$ and $P$ shown in Figures 4 and 5 respectively. The $\sigma_{slope}$ is about 0.015 in each case in these figures. The means of the 1-year correction factors serve as final correction factors between the datasets. The uncertainty of the final correction factor is determined as a combination of the standard deviation of the 1-year factors ($\sigma_{year}\sim$0.05-0.08), and the average of their $\sigma_{slope}$ values: $\sigma_{factor}=\sqrt{\sigma_{year}^2+\sigma_{slope}^2}$.

The final correction factor adjusting $K$ to $G$ is the mean of the 1-year correction factors for $K$: 

\begin{equation}
   G=0.979(\pm 0.077) *K.
\end{equation}

Before determining the correction factor adjusting $D$ to $G$, we have to determine the mean of the 1-year correction factors between $K$ and $D$. 
However, we can derive two different equations depending on that which is the independent variable. 
If the independent variable is $K$:
\begin{equation}
   D=0.925(\pm 0.055) *K. 
\end{equation}
If the independent variable is $D$: 
\begin{equation}
 K=1.085(\pm 0.073) *D.
\end{equation}

Both Equation 4 and 5 are suitable for substituting $K$ for a term containing $D$ in Equation 3. Thus, we have two ways to derive the connection between $G$ and $D$. Near $G$ the number of the Equations used for the calculation are indicated in the subscript. 

By substituting $K=D/0.925$ from Equation 4, we find  $G=0.979*D/0.925$ (and $\sigma$ is calculated by applying standard error propagation rules) then
\begin{equation}
G_{(3,4)}=1.058(\pm 0.104) *D,  
\end{equation}
where the subscripts refer to the equations used in the derivation, i.e., equations 3 and 4.

By substituting $K = 1.085 * D$ from Equation 5, we find
 $G=0.979*1.085*D$ then 
\begin{equation}
G_{(3,5)}=1.062(\pm 0.103) *D,  
\end{equation}
where the subscripts refer to the Equations 3 and 5.

Averaging the results of these two ways of calculations, Equations 6 and 7 give the following equation for the correction factor between $G$ and $D$ determined via $K$:
\begin{equation}
G_{(3,4,5)}=1.060(\pm 0.104) *D,
\end{equation}
where the subscripts refer to the Equations 3, 4 and 5.

In a similar way, we also compute the correction factor between $G$ and $D$ via $P$.
The related basic equations similar to Equations 3-5 are:
\begin{equation}
   G=1.004(\pm 0.080) *P.
\end{equation}

\begin{equation}
   D=0.909(\pm 0.052) *P 
\end{equation}

\begin{equation}
 P=1.103(\pm 0.064) *D.
\end{equation}
  
By using these equations, we derive an equation similar to Equation 8:

\begin{equation}
G_{(9,10,11)}=1.106(\pm 0.108) *D,  
\end{equation}
where the subscripts refer to the Equations 9, 10 and 11.
Now we have two equations for the correction factor between $G$ and $D$: the Equation 8 determined via $K$ and Equation 12 determined via $P$.
Averaging the Equations 8 and 12, and rounding the result to hundredths, we derive our estimation for the final correction factor between $G$ and $D$ as 

\begin{equation}
G_{(8,12)}=1.08(\pm 0.11) *D,  
\end{equation}
where the subscripts refer to the Equations 8 and 12.

\begin{figure*}

\centering
\includegraphics[width=12cm]{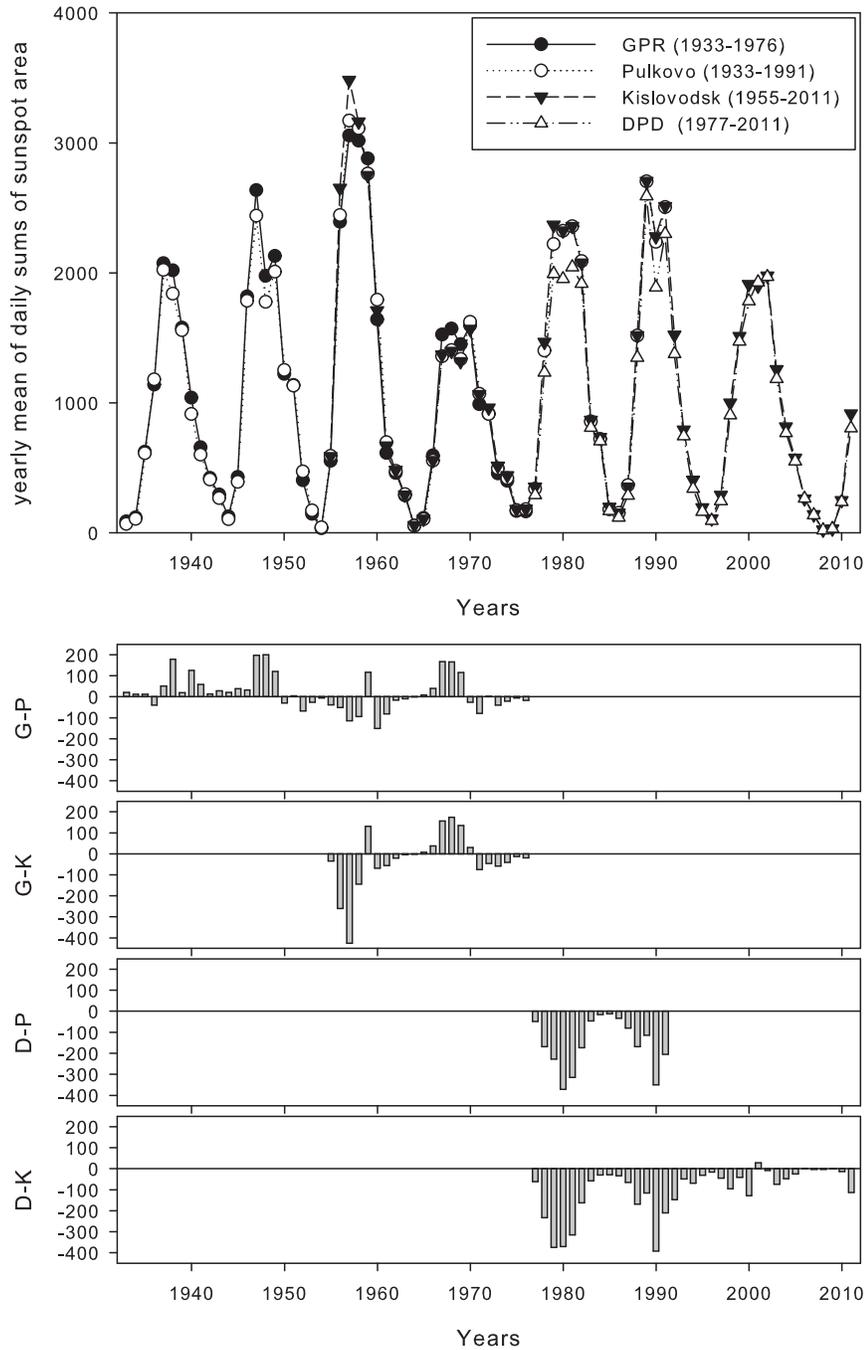}

 \caption{ Upper panel: Yearly means of daily sums of corrected sunspot area data of GPR, Pulkovo, Kislovodsk and DPD. Lower panel: Differences between the yearly means (G=GPR, P=Pulkovo, K=Kislovodsk, D=DPD).  }

 \label{fig5}

\end{figure*}
\begin{figure*}

\centering
\includegraphics[width=12cm]{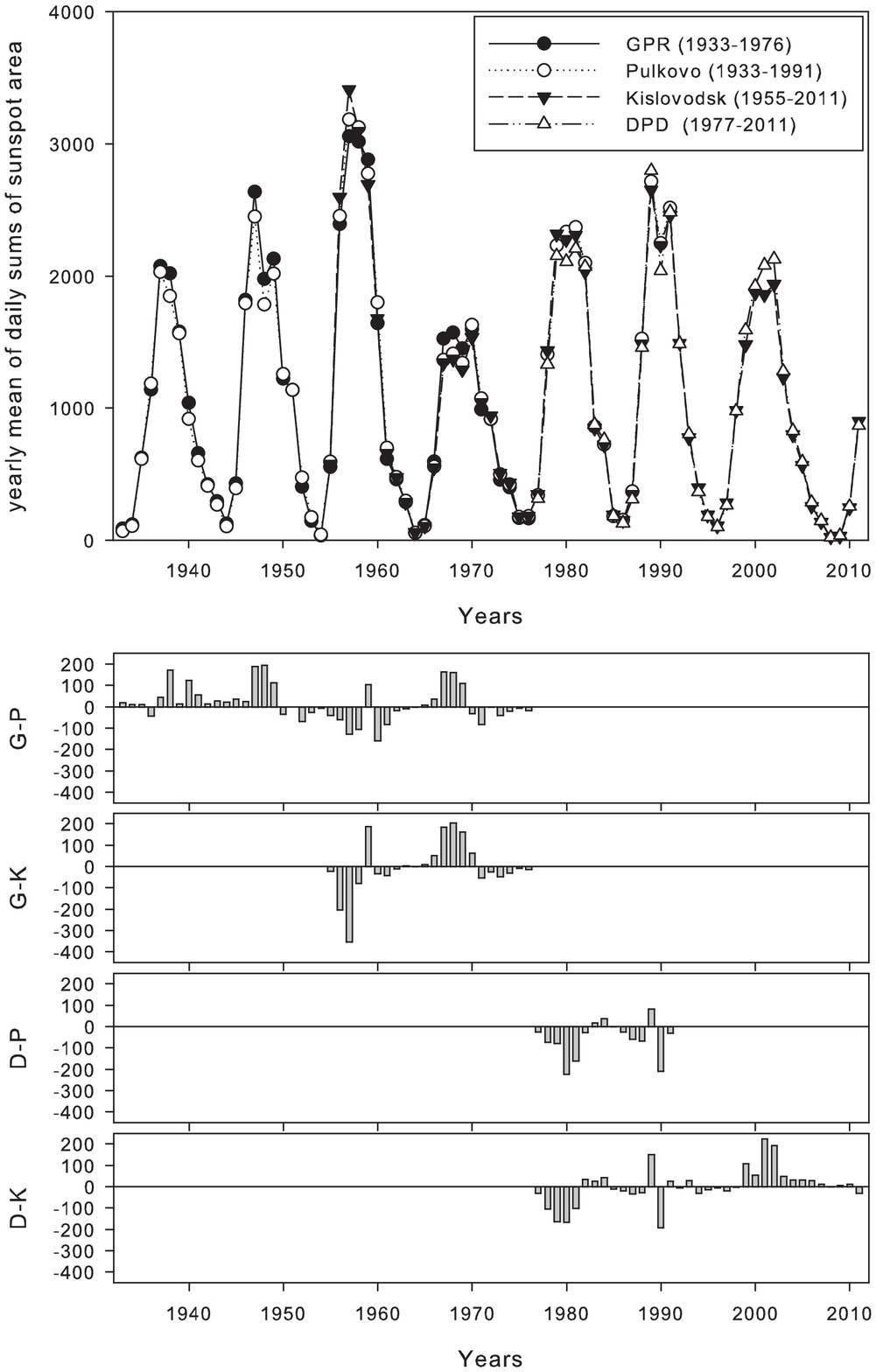}

 \caption{The same as Figure 6 but the Pulkovo, Kislovodsk and DPD data are calibrated to the level of GPR with the final calibration factors derived in this study. The Kislovodsk data are corrected to the GPR level by using Equation 3, the Pulkovo data by using Equation 9, and DPD data by using  Equation 13.}

 \label{fig7}

\end{figure*}

The Figures 6 and 7 show the yearly means of daily sums of corrected sunspot area before and after  
calibrating the Kislovodsk data by using Equation 3, the Pulkovo data by using Equation 9, and DPD data by using  Equation 13. 

\section{CONCLUSION}

   In this study, we have estimated the calibration factor between the corrected daily sunspot area of Greenwich Photoheliographic Results (GPR) and that of its continuation, the Debrecen Photoheliographic Data (DPD). We have used a method which is based on the means of the 1-year correction factors calculated for the overlapping Kislovodsk and Pulkovo databases. This method gives the following relationship: $GPR=1.08(\pm 0.11) *DPD$. We recognize that there may be problems internally within the databases which need to be investigated, especially the time-dependent variations of the datasets.
To achieve a more precise result, we will extend the study of sunspot databases in the near future. The planned comparison of data of identified sunspot groups will probably help in taking into account the short-term and long-term changes of the different sunspot databases. We aim at comparing the GPR and DPD data directly in an overlapping interval when the new volumes of DPD allow it. The revision of the data will also improve the results. 
 Currently, several international agencies are developing software to review the digital Greenwich sunspot area database, evaluating the data integrity, making corrections based on scientific analysis and group consensus, and compiling an improved version of this invaluable long term database (Willis et al. 2013a, b).  In addition, a newly digitized Greenwich database is being quality controlled to be available for comparison tests with the older version (Erwin et al. 2013). In the future, we intend to update the GPR-DPD correlation results when a more accurate Greenwich database becomes available.
Our results show that every dataset may vary in time to a smaller or larger extent. Even if the main priority of an observatory is to keep its area dataset in perfectly homogeneous state devoting large efforts to this task, it is almost impossible to avoid the effect of possible long-term changes of observational material or other observational conditions. Since the intervals of variation and stability are different in the different datasets, we conclude that the more observatories observe, measure, and publish the sunspot area data, the better for the long-term studies regardless of the observational techniques. This also underlines how important the data rescue programs are which aim at digitizing, measuring (or revising, or re-measuring with different methods), and publishing existing sunspot image and data archives. These efforts are valuable contributions to the studies of long-term variation of solar activity, and they are indispensable to achieve better results in this field.

\section*{Acknowledgments}

This work was supported by the ESA PECS project No. C98081 (T.B.). The authors are greatly indebted to a referee for suggesting substantial improvements to the presentation of the material included in the paper.

\label{lastpage}

\end{document}